\documentclass[prx,twocolumn,showpacs]{revtex4-1}
\usepackage{amsmath,amssymb}
\usepackage{bm}
\usepackage{graphicx}
\usepackage{hyperref}

\begin{document}

\title{Can the Abraham light momentum and energy in a medium \\ constitute a Lorentz four-vector?}

\author{Changbiao Wang} 
\email{changbiao\_wang@yahoo.com}
\affiliation{ShangGang Group, 70 Huntington Road, Apartment 11, New Haven, CT 06512, USA}

\begin{abstract}
By analyzing the Einstein-box thought experiment with the principle of relativity, it is shown that Abraham's light momentum and energy in a medium cannot constitute a Lorentz four-vector, and they consequentially break global momentum and energy conservation laws.  In contrast, Minkowski's momentum and energy always constitute a Lorentz four-vector no matter whether in a medium or in vacuum, and the Minkowski's momentum is the unique correct light momentum.  A momentum-associated photon mass in a medium is exposed, which explains why only the Abraham's momentum is derived in the traditional ``center-of-mass-energy'' approach.  The EM boundary-condition matching approach, combined with Einstein light-quantum hypothesis, is proposed to analyze this thought experiment, and it is found for the first time that only from Maxwell equations without resort to the relativity, the correctness of light momentum definitions cannot be identified.  Optical pulling effect is studied as well.
\end{abstract}

\maketitle

\section{Introduction}

The momentum of light in a medium is a fundamental question and has kept attracting extensive interest \cite{r1,r2,r3,r4,r5,r6,r7,r8,r9,r10,r11,r12,r13,r14,r15}.  There are different ways to define the light momentum, which all have their own merits \cite{r1,r4,r10}.  However in this paper, the light momentum is defined as single photon's momentum or electromagnetic (EM) momentum.  According to this definition, the single photon's momentum and energy are the direct result of Einstein light-quantum hypothesis of EM momentum and energy.  
 
The plane-wave model is the simplest physical model for studying light momentum, and it can be strictly treated mathematically in the Maxwell-equation frame; however, the physical results obtained are fundamental.  For example, the Lorentz transformation of photon density in the isotropic-fluid model treated by sophisticated field theory is exactly the same as that in the plane-wave model \cite{r15}. 

As a fundamental hypothesis in the special theory of relativity, the principle of relativity requires that the laws of physics be the same in form in all inertial frames of reference.  Therefore, all inertial frames are equivalent and there is no preferred inertial frame for descriptions of physical phenomena.  For example, Maxwell equations, Fermat’s principle, and the conservation laws of global momentum and energy are all valid in any inertial frames, no matter whether the medium is moving or at rest, and no matter whether the space is partially or fully filled with a medium.

For an ideal plane wave (phase velocity equal to group velocity or energy velocity), the phase function characterizes the propagation of energy and momentum of light.  (1) The light momentum is parallel to the wave vector, and (2) the phase function is a Lorentz invariant.  As physical laws, according to the principle of relativity, the above two basic properties are valid in any inertial frames.  From this we can conclude that the correct light momentum and energy must constitute a Lorentz covariant four-vector, and the Minkowski’s momentum is the unique correct light momentum \cite{r15}.

Why should the light momentum be parallel to the wave vector?  Conceptually speaking, the direction of photon propagation is the direction of photon's momentum and energy propagation.  The plane-wave phase function defines all equi-phase planes of motion, with the wave vector as their normal vector.  From one equi-phase plane to another equi-phase plane, the path parallel to the normal vector is the shortest.  Fermat's principle indicates that, light follows the path of least time.  Thus the direction of photon propagation must be parallel to the wave vector, and so must the light momentum.  The phase function is Lorentz symmetric, namely it has exactly the same form in all inertial frames.  Consequently, this property of light momentum must be valid in all inertial frames.

In a recent Letter by Barnett \cite{r7}, a widely-accepted total-momentum model is analyzed for resolution of the Abraham-Minkowski debate, where both Abraham's and Minkowski's momentums are shown to be the correct light momentums, with the former being the kinetic momentum and the latter the canonical momentum.

In this paper, by analyzing the total momentum model \cite{r7} with the principle of relativity for a dielectric-medium Einstein-box thought experiment (also called ``Balazs thought experiment'') \cite{r11,r12},  it is shown that the Abraham's momentum and energy in a medium cannot constitute a Lorentz four-vector, and they consequentially break the global momentum and energy conservation laws when a photon enters the medium box from vacuum.  Accordingly, it is not justifiable to take the Abraham's momentum as the correct light momentum.  In contrast, Minkowski's momentum and energy always constitute a Lorentz four-vector no matter whether in a medium or in vacuum; thus the Minkowski's momentum is the unique correct light momentum.  A new kind of mass, momentum-associated photon mass in a medium is exposed, which explains why the Abraham's momentum is obtained in the traditional ``center-of-mass-energy'' analysis of this thought experiment \cite{r7}.  The EM boundary-condition matching approach, combined with Einstein light-quantum hypothesis, is proposed to analyze this thought experiment, and it is found for the first time that only from Maxwell equations without resort to the principle of relativity, the correctness of light momentum definitions cannot be identified.  Optical pulling effect is studied as well.

\section{Abraham's Photon Momentum Breaking the Global Momentum and Energy Conservation Laws}

As a physical law, according to the principle of relativity, the total momentum model \cite{r7} should be valid in any inertial frames.  When applying this model to the Einstein-box thought experiment, with a slightly different process from that in \cite{r7}, where a single photon has entered a block of transparent medium with a dimension much larger than the photon's wavelength, observed in the medium-rest frame the total momentum is equal to Abraham's photon momentum, because the medium kinetic momentum is zero.  However, as shown in  Appendix \ref{appA}, the Abraham's photon momentum cannot be used to constitute a Lorentz covariant momentum-energy four-vector; consequently, the total momentum cannot either.  Similarly, observed in the lab frame, the momentum and energy of the medium box independently constitute a four-vector while the Abraham's photon momentum and energy do not.  Thus the four-vector Lorentz covariance of the total momentum and energy is questionable.  \\
\indent It is a well-known postulate that the total (global) momentum and energy are conservative for an isolated physical system \cite{r13}, which is valid in all inertial frames.  The photon and the medium box form an isolated system in the Einstein-box thought experiment.  Based on the Lorentz property of Abraham's photon momentum, we have shown in above that the total momentum and energy cannot constitute a Lorentz four-vector after the photon has entered the medium box.  In fact, by taking advantage of the initial conditions, we can further show that the Abraham's photon momentum in a medium is not compatible with the momentum and energy conservation laws.  The derivations are given below. \vspace{3mm} \\
\indent  (1) Suppose that before the photon enters the medium box, the photon initially is located far away from the medium box in vacuum.  Thus initially the photon's Abraham (= Minkowski) momentum and energy $(\textbf{p}_{\mathrm{Abr}}, E_{\mathrm{Abr}}/c)_{\mathrm{before}}$ constitute a Lorentz four-vector.  \vspace{2mm} \\
\indent (2) The medium box is made up of massive particles, and its kinetic momentum and energy $(\textbf{p}^{\mathrm{med}}_{\mathrm{kin}}, E^{\mathrm{med}}/c)$  constitute a Lorentz four-vector no matter \emph{before or after} the photon enters the medium box.   \vspace{2mm} \\
\indent (3) From (1) and (2), initially the total momentum and energy constitute a four-vector, namely
\begin{align}
(\textbf{p}_{\mathrm{total}}, E_{\mathrm{total}}/c)_{\mathrm{before}}& =(\textbf{p}_{\mathrm{Abr}}, E_{\mathrm{Abr}}/c)_{ \mathrm{before}}  \nonumber \\
&+(\textbf{p}^{\mathrm{med}}_{\mathrm{kin}}, E^{\mathrm{med}}/c)_{\mathrm{before}}  \nonumber
\end{align}
 is a four-vector.  \vspace{2mm} \\
\indent (4) According to the momentum and energy conservation laws, the total momentums and energies are equal \emph{before} and \emph{after} the photon enters the box, namely 
\begin{equation}
(\textbf{p}_{\mathrm{total}}, E_{\mathrm{total}}/c)_{\mathrm{after}}=(\textbf{p}_{\mathrm{total}}, E_{\mathrm{total}}/c)_{\mathrm{before}}. \nonumber
\end{equation}
 From (3), we know that $(\textbf{p}_{\mathrm{total}}, E_{\mathrm{total}}/c)_{\mathrm{before}}$   is a four-vector, and thus
 \begin{align}
(\textbf{p}_{\mathrm{total}}, E_{\mathrm{total}}/c)_{\mathrm{after}}&=(\textbf{p}^{\mathrm{med}}_{\mathrm{kin}}, E^{\mathrm{med}}/c)_{\mathrm{after}}  \nonumber \\
&+(\textbf{p}_{\mathrm{Abr}},~ E_{\mathrm{Abr}}/c)_{ \mathrm{after}}  \nonumber
\end{align}
also is a four-vector.  Further, because $(\textbf{p}^{\mathrm{med}}_{\mathrm{kin}}, E^{\mathrm{med}}/c)_{\mathrm{after}}$ ~ is a four-vector resulting from (2), ~~$(\textbf{p}_{\mathrm{Abr}}, E_{\mathrm{Abr}}/c)_{ \mathrm{after}}$  \emph{must} be a four-vector.  However $(\textbf{p}_{\mathrm{Abr}}, E_{\mathrm{Abr}}/c)_{ \mathrm{after}}$  \emph{cannot} be a four-vector according to the principle of relativity [confer Eq.\ (\ref{eqA3}) in Appendix \ref{appA}].  Thus we conclude that the Abraham's photon momentum contradicts the momentum and energy conservation laws in the principle-of-relativity frame, which means that the Abraham's photon momentum cannot make the conservation laws holding in all inertial frames --- the direct physical consequences of Abraham's light momentum. \vspace{3mm} \\
\indent From the above relativity analysis of Einstein-box thought experiment, we can see that the correct light momentum and energy must constitute a Lorentz four-vector when the global momentum and energy conservation laws are taken to be the fundamental postulates \cite{r13}, which actually is a criterion of identifying the correctness of light momentum definitions.  This conclusion is completely in agreement with that obtained from a plane wave in a moving uniform medium \cite{r15}, as stated in Sec. I. \\
\indent It is interesting to point out that, it is the Fermat's principle and the principle of relativity that require the light momentum and energy to constitute a Lorentz four-vector for a plane wave in a moving uniform medium where there is no momentum transfer taking place \cite{r15}, while it is the global momentum and energy conservation laws that require the light momentum and energy to constitute a Lorentz four-vector in the Einstein-box thought experiment where there is a momentum transfer taking place when the photon goes into a medium box.
\\
\\
\\

\section{Einstein-Box Thought Experiment Analyzed by EM Boundary-Condition Matching Approach}

It has been shown that the Abraham's light momentum and energy for a plane wave in a uniform medium is not Lorentz covariant \cite{r15}.  The plane wave is a strict solution of Maxwell equations in the macro-scale electromagnetic theory, and this solution indicates that the Minkowski's momentum density vector $\mathbf{D}\times\mathbf{B}$  and energy density $\mathbf{D}\cdot\mathbf{E}=\mathbf{B}\cdot\mathbf{H}$ constitute a Lorentz four-vector in the form of $\bar{P}^{\mu}=N^{-1}_p(\mathbf{D}\times\mathbf{B}, \mathbf{D}\cdot\mathbf{E}/c)$, with $N_p$  the ``EM-field-cell density'' or ``photon density'' in volume \cite{r15}, and $c$ the vacuum light speed.  When Einstein's light-quantum hypothesis $N^{-1}_p\mathbf{D}\cdot\mathbf{E}=\hbar\omega$  is imposed, $\bar{P}^{\mu}$  is restored to a single photon's momentum-energy four-vector, namely $N^{-1}_p(\mathbf{D}\times\mathbf{B}, \mathbf{D}\cdot\mathbf{E}/c)=(\hbar n_d\mathbf{k}, \hbar\omega/c)$, with  $n_d\mathbf{k}$  the wave vector; thus $\mathbf{D}\times\mathbf{B}$   denotes the unique correct light momentum.  For the plane wave in a uniform medium, $\mathbf{D}\times\mathbf{B}=|n_d(\mathbf{D}\cdot\mathbf{E})/c|\mathbf{\hat{n}}$ holds in all inertial frames, where $n_d$ is the refractive index and  $\mathbf{\hat{n}}$  is the unit wave vector \cite{r15}.

Now let us apply the Minkowski's momentum to analysis of a plane-wave light pulse perpendicularly incident on the above transparent medium box without any reflection \cite{r12,r14}.  The pulse space length is assumed to be much larger than the wavelength but less than the box length.  To eliminate any reflection, the wave-impedance matching must be reached between vacuum and the medium \cite{r12}, namely the wave impedance $(\mu/\epsilon)^{1/2}$  with $\mu=\mathbf{B}/\mathbf{H}$  and $\epsilon=\mathbf{D}/\mathbf{E}$  is continuous on the boundary (confer Fig. \ref{fig1} in Appendix \ref{appB}). 

Since there is no reflection, there is no energy accumulation in the sense of time average.  Thus ``no-reflection'' can be expressed as ``equal energy flux density'' on the both sides of the vacuum-medium interface inside the light pulse, given by 
\begin{equation*}
(\mathbf{D}\cdot\mathbf{E})_{vac}c=(\mathbf{D}\cdot\mathbf{E})_{med}\left( \frac{c}{n_d} \right), ~~~\mathrm{or}
\end{equation*} 
\begin{equation} 
(\mathbf{D}\cdot\mathbf{E})_{med}=n_d(\mathbf{D}\cdot\mathbf{E})_{vac}.~~~~~~~~~~~~~~
\label{eq1}
\end{equation}

The above Eq.\ (\ref{eq1}) is indeed equivalent to the wave-impedance matching condition $(\mu/\epsilon)_{med}=(\mu/\epsilon)_{vac}$ when the perpendicularly-incident plane-wave boundary condition $(\mathbf{E}\cdot\mathbf{E})_{med}=(\mathbf{E}\cdot\mathbf{E})_{vac}$  is considered, because in such a case we have
\begin{equation*}
\begin{array} {ll}
(\mathbf{D}\cdot\mathbf{E})_{med}&=\epsilon_{med}(\mathbf{E}\cdot\mathbf{E})_{med}   \vspace{2mm}  \\ 
&=(\epsilon/\mu)_{med}^{1/2}(\epsilon\mu)_{med}^{1/2}(\mathbf{E}\cdot\mathbf{E})_{med} \vspace{2mm}  \\
&=(\epsilon/\mu)_{vac}^{1/2}[n_d(\epsilon\mu)_{vac}^{1/2}](\mathbf{E}\cdot\mathbf{E})_{vac} \vspace{2mm}  \\
&=n_d(\mathbf{D}\cdot\mathbf{E})_{vac}, 
\end{array}
\end{equation*}
namely Eq.\ (\ref{eq1}), where $(\epsilon\mu)^{1/2}_{med}=n_d(\epsilon\mu)^{1/2}_{vac}$  is employed.

From Eq.\ (\ref{eq1}), we have 
\begin{equation*}
\begin{array}{ll}
(\mathbf{D}\times\mathbf{B})_{med}&=[n_d(\mathbf{D}\cdot\mathbf{E})_{med}/c]\mathbf{\hat{n}} \nonumber   \vspace{2mm}  \\  
&=[n_d^2 (\mathbf{D}\cdot\mathbf{E})_{vac}/c]\mathbf{\hat{n}}  \nonumber   \vspace{2mm}  \\ 
&=n_d^2 (\mathbf{D}\times\mathbf{B})_{vac}, 
\end{array}
\end{equation*}
namely 
\begin{equation}
(\mathbf{D}\times\mathbf{B})_{med}=n_d^2 (\mathbf{D}\times\mathbf{B})_{vac}.~~~~~
\label{eq2}
\end{equation}

The momentum flux density in the medium is $(\mathbf{D}\times\mathbf{B})_{med}\cdot(c/n_d)\mathbf{\hat{n}}$, while the momentum flux density in the vacuum is $(\mathbf{D}\times\mathbf{B})_{vac}\cdot c\mathbf{\hat{n}}$.  Thus from Eq.\ (\ref{eq2}) we have 
\begin{equation}
[(\mathbf{D}\times\mathbf{B})_{med}\cdot(c/n_d)\mathbf{\hat{n}}]=n_d[(\mathbf{D}\times\mathbf{B})_{vac}\cdot c\mathbf{\hat{n}}].
\label{eq3}
\end{equation}

Eq.\ (\ref{eq3}) tells us that, after the Minkowski's EM momentum (in unit area and unit time) flows into the medium from vacuum, the momentum grows by $n_d$  times.  To keep the total momentum unchanged, there must be a pulling force acting on the medium when the plane-wave light pulse goes into the medium box (see Appendix \ref{appB}), which is the result from macro-electromagnetic theory based on the assumption of ``no reflection''.  This pulling force can be qualitatively explained as the Lorentz force produced by the interaction of the dielectric bound current with the incident light pulse \cite{r12}. 

Now let us examine the result from Einstein light-quantum theory.  The photon energy (frequency) is supposed to be the same no matter whether in vacuum or in a medium.  Einstein light-quantum hypothesis requires that $(\mathbf{D}\cdot\mathbf{E})_{med}=N_p^{(med)} \hbar\omega$ and  $(\mathbf{D}\cdot\mathbf{E})_{vac}=N_p^{(vac)} \hbar\omega$, with $N_p^{(med)}$  the photon density in medium and $N_p^{(vac)}$  the photon density in vacuum.  Inserting them into Eq.\ (\ref{eq1}), we have 
\begin{equation}
N_p^{(med)}=n_d N_p^{(vac)}.
\label{eq4}
\end{equation}

Supposing that $\mathbf{p}^{(\mathrm{Min})}_{med-photon}$  and $\textbf{p}_{vac-photon}=(\hbar\omega/c)\mathbf{\hat{n}}$  are the photon momentums in medium and in vacuum respectively, from the definition of momentum density we have 
\begin{equation}
\begin{array} {ll}
(\mathbf{D}\times\mathbf{B})_{med}&=N_p^{(med)}\mathbf{p}_{med-photon}^{(\mathrm{Min})}   \vspace{2mm}  \\ 
(\mathbf{D}\times\mathbf{B})_{vac}&=N_p^{(vac)}~\mathbf{p}_{vac-photon}
\end{array}
\label{eq5}
\end{equation}

Inserting Eqs.\ (\ref{eq4}) and (\ref{eq5}) into Eq.\ (\ref{eq3}), we have the photon momentum in the medium, given by
\begin{equation*}
\mathbf{p}^{(\mathrm{Min})}_{med-photon}=n_d\hspace{0.05em}\textbf{p}_{vac-photon}, ~~~~\mathrm{or} 
\end{equation*}
\begin{equation}
\mathbf{p}^{(\mathrm{Min})}_{med-photon}=\frac{n_d \hbar\omega}{c} \mathbf{\hat{n}}.~~~~~~~~~~~~~~~
\label{eq6}
\end{equation}

From Eq.\ (\ref{eq6}) we can see that when a single photon goes into the medium box, the medium box also gets a pulling force to keep the total momentum unchanged.

From Eq.\ (\ref{eq3}) and Eq.\ (\ref{eq6}) we find that a light pulse and a single photon in the medium-box thought experiment both have the pulling effect.  How do we explain the fiber recoiling experiment then \cite{r5}?  The recoiling could be resulting from the transverse radiation force because of an azimuthal asymmetry of refractive index in the fiber \cite{r9}.

\section{Implicit Assumption of Photon's Mass in the Traditional ``Center of Mass-Energy'' Argument}

It is worthwhile to point out that, the widely-recognized ``center of mass-energy'' argument for Abraham's photon momentum \cite{r7} is questionable.  As shown in Appendix \ref{appC}, this argument neglects the difference between the ``momentum-associated mass'' and ``energy-associated mass'' for a photon in a medium.  Specifically speaking, this argument has implicitly assumed that the relation between photon's ``momentum-associated'' mass and its momentum in a dielectric is the same as that in vacuum.  The photon momentum-associated mass in vacuum, formulated by $\hbar\omega/c^2$, is derived from ``vacuum'' Einstein-box thought experiment \cite{r16,r17}, and whether the formulation still holds in a ``dielectric'' remains to be confirmed.  Now that this assumption has already resulted in contradictions with the covariance of relativity, the justification of the assumption should be re-considered.

\section{Conclusions}

In summary, by analysis of the total momentum model \cite{r7} with the principle of relativity for a medium Einstein-box thought experiment, we have shown that the Abraham's light momentum and energy in a medium do not constitute a Lorentz four-vector, and they break the global momentum-energy conservation law; accordingly, it is not justifiable to take the Abraham's momentum as the correct light momentum.  In contrast, Minkowski's momentum and energy always constitute a four-vector no matter whether in a medium or vacuum, and the Minkowski's momentum is the unique correct light momentum.  This result of the relativity principle is important, because only based on the Maxwell equations one cannot judge which formulation of light momentum is correct.  For example, inserting $(\mathbf{D}\times\mathbf{B})_{med}=n^2_d(\mathbf{E}\times\mathbf{H}/c^2)_{med}$ and  $(\mathbf{D}\times\mathbf{B})_{vac}=(\mathbf{E}\times\mathbf{H}/c^2)_{vac}$   into Eq.\ (\ref{eq3}), we directly obtain the conversion equation for Abraham's momentum flux density from vacuum to medium, given by
\begin{align}
[(\mathbf{E}\times\mathbf{H}/c^2)_{med}\cdot &(c/n_d)\mathbf{\hat{n}}]  \nonumber \\
&=\frac{1}{n_d}[(\mathbf{E}\times\mathbf{H}/c^2)_{vac}\cdot c\mathbf{\hat{n}}],
\label{eq7}
\end{align}
and ~~similarly, ~~inserting  ~~$(\mathbf{E}\times\mathbf{H}/c^2)_{med}=N_p^{(med)}\mathbf{p}^{(\mathrm{Abr})}_{med-photon}$,  $(\mathbf{E}\times\mathbf{H}/c^2)_{vac}=N_p^{(vac)}\textbf{p}_{vac-photon}$, and $N^{(med)}_p=n_d N^{(vac)}_p$  into above Eq.\ (\ref{eq7}), we have the Abraham's photon momentum in medium, given by 
\begin{equation*}
\mathbf{p}^{(\mathrm{Abr})}_{med-photon}=\frac{1}{n_d} \textbf{p}_{vac-photon}, ~~~\mathrm{or}
\end{equation*}
\begin{equation}
\mathbf{p}^{(\mathrm{Abr})}_{med-photon}=\frac{\hbar\omega}{n_d c}\mathbf{\hat{n}}.~~~~~~~~~~~~~~~~~
\label{eq8}
\end{equation}
Thus in the Maxwell-equation frame, the medium Einstein-box thought experiment supports both light momentum formulations, instead of just Abraham's \cite{r7}.  However, the two formulations cannot be ``both correct''; otherwise it is not determinate whether the medium box gets a pulling force or a pushing force when a specific photon goes into the medium from vacuum.  In other words, without resort to the principle of relativity, this thought experiment cannot be used to identify the correctness of light momentum definitions.


\appendix
\section{Lorentz property of the Abraham's photon momentum and energy}
\label{appA}
In this Appendix, by analysis of the Lorentz property of the total momentum model in the dielectric-medium Einstein-box thought experiment \cite{r7}, a specific proof is given of why the Abraham's photon momentum and energy cannot constitute a Lorentz four-vector.

According to the total-momentum model \cite{r7}, the total momentum and the total energy are assumed to constitute a momentum-energy four-vector.  

Suppose that the total momentum $\mathbf{p}_{\mathrm{total}}$  and the total energy $E_{\mathrm{total}}$  in the lab frame are written as 
\begin{equation}
\left\{
\begin{array}{ll}
&\mathbf{p}_{\mathrm{total}}=\mathbf{p}_{\mathrm{kin}}^{\mathrm{med}}+\textbf{p}_{\mathrm{Abr}}, \vspace{2mm} \\ &E_{\mathrm{total}}=E^{\mathrm{med}}+E_{\mathrm{Abr}},  \vspace{2mm} \\
&P^{\mu}=(\mathbf{p}_{\mathrm{total}}, E_{\mathrm{total}}/c)~~\textrm{a four-vector},
\end{array} \right.
\label{eqA1}
\end{equation}
where $\mathbf{p}_{\mathrm{kin}}^{\mathrm{med}}$  and $E^{\mathrm{med}}$  are, respectively, the medium-box kinetic momentum and energy, while $\textbf{p}_{\mathrm{Abr}}$  and $E_{\mathrm{Abr}}$  are, respectively, the Abraham's photon momentum and energy.

After the single photon has entered the Einstein's medium box, according to the principle of relativity (\emph{the laws of physics are the same in form in all inertial frames}), the total momentum and energy in the medium-rest frame can be written as
\begin{equation}
\left\{
\begin{array}{ll}
&\mathbf{p}'_{\mathrm{total}}=\mathbf{p}'^{\mathrm{med}}_{\mathrm{kin}}+\textbf{p}'_{\mathrm{Abr}}, \vspace{2mm} \\ &E'_{\mathrm{total}}=E'^{\mathrm{med}}+E'_{\mathrm{Abr}},  \vspace{2mm} \\
&P'^{\mu}=(\mathbf{p}'_{\mathrm{total}}, E'_{\mathrm{total}}/c)~~\textrm{a four-vector},
\end{array} \right.
\label{eqA2}
\end{equation}
where $\mathbf{p}'^{\mathrm{med}}_{\mathrm{kin}}$  and $E'^{\mathrm{med}}$  are, respectively, the medium-box kinetic momentum and energy, while $\textbf{p}'_{\mathrm{Abr}}$ and $E'_{\mathrm{Abr}}$ are, respectively, the Abraham's photon momentum and energy.  

Since $ P^{\mu}$ is assumed to be a Lorentz four-vector, $P^{\mu}=(\mathbf{p}_{\mathrm{total}}, E_{\mathrm{total}}/c)$  can be obtained from $P'^{\mu}=(\mathbf{p}'_{\mathrm{total}}, E'_{\mathrm{total}}/c)$ by Lorentz transformation.  
 
Now let us examine whether the total momentum $\mathbf{p}_{\mathrm{total}}=\mathbf{p}_{\mathrm{kin}}^{\mathrm{med}}+\textbf{p}_{\mathrm{Abr}}$ \cite{r7} can really make $(\mathbf{p}_{\mathrm{total}}, E_{\mathrm{total}}/c)$  become a Lorentz four-vector.

In the medium-rest frame, the medium kinetic momentum is equal to zero, namely $\mathbf{p}'^{\mathrm{med}}_{\mathrm{kin}}=0$, and the total momentum is reduced to $\mathbf{p}'_{\mathrm{total}}=\mathbf{p}'^{\mathrm{med}}_{\mathrm{kin}}
+\textbf{p}'_{\mathrm{Abr}}=\textbf{p}'_{\mathrm{Abr}}$.
\vspace{3mm} \\
\indent (i) The medium-box kinetic momentum $\mathbf{p}'^{\mathrm{med}}_{\mathrm{kin}}=0$  and its rest energy $E'^{\mathrm{med}}$  independently constitute a Lorentz four-vector, namely $(\mathbf{p}'^{\mathrm{med}}_{\mathrm{kin}}, E'^{\mathrm{med}}/c)$  is a four-vector. 
\vspace{2mm} \\
\indent (ii) The Abraham's photon momentum and energy is given by
\begin{equation}
\left( \mathbf{p}'_{\mathrm{Abr}},\frac{E'_{\mathrm{Abr}}}{c} \right)=\left( \frac{\hbar\omega'}{n'_d c}\mathbf{\hat{n}}',\frac{\hbar\omega'}{c} \right),
\label{eqA3}
\end{equation}
where $n'_d$  is the refractive index of medium, $\omega'$  is the photon's frequency, $\mathbf{\hat{n}}'$  is the unit vector of the photon's moving direction, and $\hbar$  is the Planck constant.  We have known that, the wave four-vector $K'^{\mu}=[(n'_d\omega'/c)\mathbf{\hat{n}}',\omega'/c]$ must be a Lorentz four-vector and the Planck constant $\hbar$  must be a Lorentz invariant \cite{r15}, and thus the Abraham's photon momentum and energy Eq.\ (\ref{eqA3}) cannot be a four-vector; otherwise, contradictions would result mathematically \cite{r18}.
\vspace{3mm} \\
\indent From (i) and (ii) we conclude that the total momentum and energy  $(\mathbf{p}'_{\mathrm{total}}, E'_{\mathrm{total}}/c)$, which are the combinations of two parts respectively, \emph{cannot} be a Lorentz four-vector. 

If $(\mathbf{p}'_{\mathrm{total}}, E'_{\mathrm{total}}/c)$  is \emph{not} a Lorentz four-vector observed in one inertial frame, then it is \emph{never} a Lorentz four-vector observed in any inertial frames.  

The above reasoning is based on the following facts: 
\begin{itemize}
\item General math results.  (a) If $A^{\mu}$  and $B^{\mu}$  are both Lorentz four-vectors, then $A^{\mu}\pm B^{\mu}$  must be Lorentz four-vectors.  (b) If $A^{\mu}$  is a known Lorentz four-vector in one inertial frame, then it is always a Lorentz four-vector observed in any inertial frames.
\item  In the medium Einstein-box thought experiment, like a massive particle the medium-box kinetic momentum and energy independently constitute a Lorentz four-vector because the medium box is made up of massive particles, of which each has a kinetic momentum-energy four-vector.
\end{itemize}

In summary, when applying the principle of relativity to the total momentum model \cite{r7} for the dielectric-medium Einstein-box thought experiment, we obtain the following conclusion.  After the single photon has entered the Einstein's medium box, observed in any inertial frames, 
\begin{itemize}
\item the total momentum and energy is the combination of the medium-box and Abraham's momentums and energies, namely   $(\textbf{p}_{\mathrm{total}}, E_{\mathrm{total}}/c)=(\textbf{p}^{\mathrm{med}}_{\mathrm{kin}}, E^{\mathrm{med}}/c)+(\textbf{p}_{\mathrm{Abr}}, E_{\mathrm{Abr}}/c)$; 
\item the medium-box momentum and energy $(\textbf{p}^{\mathrm{med}}_{\mathrm{kin}}, E^{\mathrm{med}}/c)$  \emph{must} be a Lorentz four-vector;
\item the Abraham's photon momentum and energy $(\textbf{p}_{\mathrm{Abr}}, E_{\mathrm{Abr}}/c)$  \emph{cannot} be a Lorentz four-vector.
\end{itemize}
Therefore, the total momentum and energy $(\textbf{p}_{\mathrm{total}}, E_{\mathrm{total}}/c)$  is not a Lorentz four-vector.  \vspace{3mm} \\
\indent One might argue that instead of Eq.\ (\ref{eqA3}) the photon's momentum and energy should be   $[\mathbf{\hat{n}}'\hbar\omega'/(n'_dc),\hbar\omega'/(n'_dc)]$, where the photon energy is replaced by Chu's energy $\hbar\omega'/n'_d$ \cite{r13}.  However it can be shown that this kind of momentum and energy also contradicts the wave four-vector $K^{\mu}$.  That is because the Lorentz transformation of frequency (Doppler formula) and the transformation of refractive index are already defined by $K^{\mu}$ \cite{r15}.   Defining a four-momentum will introduce new Lorentz transformations of the frequency and refractive index, which contradict the formers unless the four-momentum is compatible with $K^{\mu}$.  The Minkowski's four-momentum $\hbar K^{\mu}$, no matter whether in a medium or in vacuum, is only different from $K^{\mu}$   by a Planck constant $\hbar$, which is a Lorentz scalar \cite{r15}, and it is completely compatible with $K^{\mu}$.  Thus only the Minkowski's momentum is the unique correct light momentum in the Einstein-box thought experiment.

One might question whether (a) the medium-box kinetic momentum $\mathbf{p}^{\mathrm{med}}_{\mathrm{kin}}$ and energy $E^{\mathrm{med}}$  can really independently constitute a Lorentz four-vector and (b) there is any medium-rest frame, because \emph{there must be relative motions between the elements of medium (fluid)}, which, even if quite small, could not be ignored in the sense of strict relativity. 
In fact, even if there are relative motions between the elements of the dielectric medium, the medium-box kinetic momentum and energy also independently constitute a four-vector, which is elucidated below.

According to the total momentum model (see Eq.\ (7) of Ref. \cite{r7}), the total momentum and energy are given by $\mathbf{p}_{\mathrm{total}}=\mathbf{p}_{\mathrm{kin}}^{\mathrm{med}}+\textbf{p}_{\mathrm{Abr}}$  and $E_{\mathrm{total}}=E^{\mathrm{med}}+E_{\mathrm{Abr}}$, which are conservative.  $\mathbf{p}_{\mathrm{kin}}^{\mathrm{med}}$  and $E^{\mathrm{med}}$ denote the \emph{medium} kinetic momentum and energy, and they are only contributed by all the massive particles of which the dielectric medium is made up, while $\textbf{p}_{\mathrm{Abr}}$ and $E_{\mathrm{Abr}}$  denote the \emph{EM} kinetic momentum and energy and they are only contributed by all EM fields or waves.  

One essential difference between massive particles and photons is that any massive particle has its four-velocity defined by $d(\mathbf{x},ct)/d\tau$  with $d\tau$ its proper time, while the photon does not \cite{r15}.  Because the medium box is made up of massive particles and each of the particles has a four-velocity, no matter whether there are any relative motions between the particles, the medium-box total kinetic momentum $\mathbf{p}_{\mathrm{kin}}^{\mathrm{med}}$  and energy $E^{\mathrm{med}}$  should constitute a four-vector.  For a better understanding, specific calculations are given below.  
 
Suppose that observed in the lab frame, the four-velocity of a massive particle is given by $dX^{\mu}_i/d\tau$, and the medium-box total kinetic momentum-energy four-vector can be written as 
\begin{equation}
(P^{\mu})^{\mathrm{med}}_{\mathrm{kin}}=\sum m_{0i}\frac{dX^{\mu}_i}{d\tau_i}=\left(\mathbf{p}_{\mathrm{kin}}^{\mathrm{med}},\frac{E^{\mathrm{med}}}{c} \right),
\label{eqA4}
\end{equation}
where
\begin{equation}
\mathbf{p}_{\mathrm{kin}}^{\mathrm{med}}=\sum m_{0i}\gamma_{ui}\mathbf{u}_i, ~~~~ \frac{E^{\mathrm{med}}}{c}=\sum m_{0i}\gamma_{ui}c,
\label{eqA5}
\end{equation} 
with $m_{0i}$, $\gamma_{ui}$, and  $\mathbf{u}_i$, respectively, the individual particles' rest mass, relativistic factor, and velocity.

Now we can define the moving velocity of the whole medium box with respect to the lab frame, given by \cite{r19}
\begin{equation}
\mathbf{v}=\frac{\sum m_{0i}\gamma_{ui}\mathbf{u}_i}{\sum m_{0i}\gamma_{ui}},~~~
\label{eqA6}
\end{equation}
and its relativistic factor $\gamma$  and the medium-box rest mass  $M^{\mathrm{med}}_0$, given by
\begin{equation}
\gamma=\frac{1}{\sqrt{1-\displaystyle{ \mathbf{v}^2/c^2 }}},~~~~~M^{\mathrm{med}}_0=\frac{\sum m_{0i}\gamma_{ui}}{\gamma}.
\label{eqA7}
\end{equation}

The medium-box kinetic momentum-energy four-vector now can be re-written as
\begin{equation}
(P^{\mu})^{\mathrm{med}}_{\mathrm{kin}}=\left(\mathbf{p}_{\mathrm{kin}}^{\mathrm{med}},\frac{E^{\mathrm{med}}}{c} \right)=\gamma M^{\mathrm{med}}_0 (\mathbf{v},c).
\label{eqA8}
\end{equation}

From above we can see that, (a) the medium-box kinetic momentum and energy indeed independently constitute a four-vector, and (b) there is a medium-rest frame for the box, which moves at the velocity $\mathbf{v}$ with respect to the lab frame defined by Eq.\ (\ref{eqA6}).  

If all particles could always keep the same velocity, this medium box would become a ``rigid body''; thus possibly causing the controversy of the compatibility with relativity.  However it should be emphasized that, in the uniform-medium model \cite{r15}, it is the dielectric parameters ($\mu=\mathbf{B}/\mathbf{H}$ and $\epsilon=\mathbf{D}/\mathbf{E}$) that are assumed to be real scalar constants observed in the medium-rest frame, instead of the medium being ``rigid''; thus this model is completely compatible with the relativity.  In fact, the uniform-medium model is widely used in literature \cite{r7,r20}, although all atoms or molecules in dielectric materials used as a uniform medium are always in constant motion or vibration.  Especially, the uniform-medium model is also strongly supported by the well-known relativity experiment, Fizeau running-water experiment, where the refractive index of the water in the water-rest frame is taken to be a constant \cite{r21}.

\section{Optical pulling effect in the Einstein-box thought experiment}
\label{appB}

\begin{figure}
\begin{center}
\includegraphics[bb=1.1in 4.6in 7.4in 10.1in, scale=0.55]{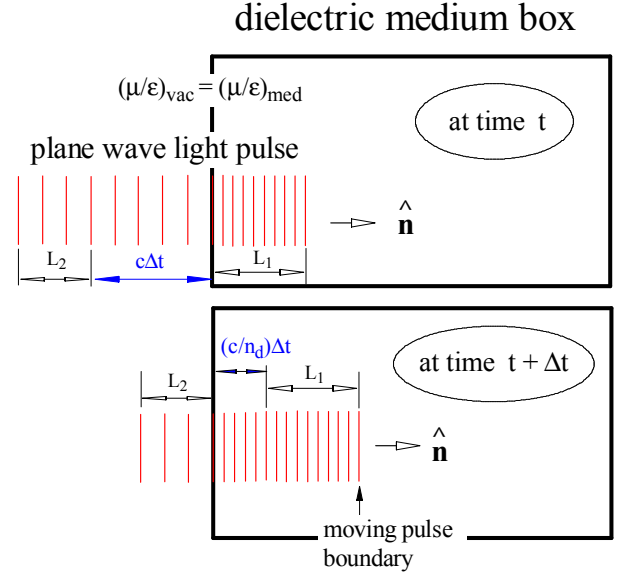}
\end{center}
\caption{Medium Einstein-box thought experiment for a light pulse, analyzed by EM boundary-condition matching approach.  A plane-wave light pulse is perpendicularly incident on the medium box without any reflection.  From time $t$  to  $t+\Delta t$, the parts of the pulse in the box with the length $L_1$ are exactly the same, and the parts in vacuum with the length $L_2$ are also the same.  The momentum difference on unit cross-section area for the pulse at $t+\Delta t$  and $t$  is only caused by the two parts with the lengths $(c/n_d)\Delta t$  and  $c\Delta t$, which is given by  $\Delta p=|(\mathbf{D}\times\mathbf{B})_{med}|(c/n_d)\Delta t-|(\mathbf{D}\times\mathbf{B})_{vac}|c\Delta t$, and an equal but different sign quantity of momentum is transferred to the medium box to keep the total momentum unchanged.  From this we obtain the force acting on the box, given by $f^{(\mathrm{Min})}=-\Delta p/\Delta t$, namely Eq.\ (\ref{eqB3}).  In addition, on the ``moving pulse boundary'' (leading or trailing edge) within the box, the EM fields are continuous because the dielectrics on the both sides of the ``moving boundary'' are exactly the same. }
\label{fig1}
\end{figure}

In the medium Einstein-box thought experiment for a light pulse, as shown in Fig.\ (\ref{fig1}), the pulling force per unit cross-section area acting on the medium box can be directly obtained from the EM boundary conditions of ``no reflection'', as shown below. 

The momentum flowing through the inner medium surface per unit area and time, observed in the ``instant medium-rest frame'', is given by 
\begin{equation}
(\mathbf{D}\times\mathbf{B})_{med}\cdot\left( \frac{c}{n_d}\mathbf{\hat{n}}\right)=(\mathbf{D}\cdot\mathbf{E})_{med},
\label{eqB1}
\end{equation}
and the momentum flowing through the inner vacuum surface per unit area and time is given by 
\begin{equation}
(\mathbf{D}\times\mathbf{B})_{vac}\cdot c\mathbf{\hat{n}}=(\mathbf{D}\cdot\mathbf{E})_{vac},
\label{eqB2}
\end{equation}
where $\mathbf{D}\times\mathbf{B}$  is the relativity-legitimate Minkowski's momentum density, and $(c/n_d)\mathbf{\hat{n}}$  and $c\mathbf{\hat{n}}$  are, respectively, the propagation velocities of EM momentum and energy in the medium and vacuum.  The ``instant medium-rest frame'' means the frame in which the medium is at rest from time $t$  to  $t+\Delta t$.

Considering $(\mathbf{D}\cdot\mathbf{E})_{med}=n_d(\mathbf{D}\cdot\mathbf{E})_{vac}$  given by Eq.\ (\ref{eq1}), which results from the electromagnetic boundary conditions, we obtain Eq.\ (\ref{eqB1})$-$Eq.\ (\ref{eqB2}) $=(\mathbf{D}\cdot\mathbf{E})_{med}-(\mathbf{D}\cdot\mathbf{E})_{vac}=(n_d-1)(\mathbf{D}\cdot\mathbf{E})_{vac}$, which is the momentum gained by the light pulse in unit cross-section area and unit time.  From this we directly obtain the Minkowski's force acting on the box, as shown in Fig.\ (\ref{fig1}), given by 
\begin{equation}
\mathbf{f}^{(\mathrm{Min})}=(1-n_d)(\mathbf{D}\cdot\mathbf{E})_{vac}\mathbf{\hat{n}}~\mathrm{[N/m^2]},
\label{eqB3}
\end{equation}
where $1-n_d<0$  means that the force direction is opposite to the direction of wave propagation, namely a \emph{pulling} force.

For a plane wave, after taking time average the pulling force is given by 
\begin{equation}
<\mathbf{f}^{(\mathrm{Min})}>=\frac{1}{2} (1-n_d)\epsilon_0\mathbf{E}^2_{vac-\mathrm{max}}\mathbf{\hat{n}},
\label{eqB4}
\end{equation}
where  $\epsilon_0=\epsilon_{vac}$, and $|\mathbf{E}_{vac-\mathrm{max}}|$  is the plane-wave electric field amplitude in vacuum.

Light-quantizing Eq.\ (\ref{eqB3}) by $(\mathbf{D}\cdot\mathbf{E})_{vac}=N^{(vac)}_p\hbar\omega$  and considering that $N^{(vac)}_p c$  is the photon number flux density (photon number through unit cross-section area in unit time in vacuum), we obtain the transferred momentum from a single photon to the medium when the photon goes into the box, given by
\begin{equation}
\mathbf{p}^{(\mathrm{Min})}_{\mathrm{transfer-to-box}}=\frac{\mathbf{f}^{(\mathrm{Min})}}{N^{(vac)}_p c}=(1-n_d)\frac{\hbar\omega}{c}\mathbf{\hat{n}}.
\label{eqB5}
\end{equation}

It should be indicated that the pulling force Eq.\ (\ref{eqB3}) is obtained without any ambiguity based on the momentum conservation law; however, some ambiguity will show up if using the surface bound current $\mathbf{J}_{bound}$  and the magnetic field $\mathbf{B}$ to calculate the force by $\mathbf{J}_{bound}\times\mathbf{B}$, because $\mathbf{B}$ is not continuous on the vacuum-medium interface \cite{r12}.

One might argue that the leading edge of the light pulse would also produce a force to cancel out the pulling force resulting from the momentum transfer on the vacuum-medium interface so that no net momentum transfer would take place \cite{r12}.  However, such an argument does not seem consistent with the dielectric property of an isotropic uniform medium.
 
An ideal isotropic uniform medium has no dispersion and losses; accordingly, any part of the pulse within the medium always keeps the same shape and the same wave momentum during propagation within the medium, as illustrated in Fig.\ (\ref{fig1}).  Thus there is no additional momentum transfer happening except for on the vacuum-medium interface.

In calculations of the Lorentz force caused by polarization and magnetization, how to appropriately approximate a light pulse is tricky.  As shown in Fig.\ (\ref{fig1}), the basic physical condition, which the pulse is required to satisfy, is the ``moving boundary condition'', namely the EM fields $\mathbf{E}$ and $\mathbf{B}$ on the leading and trailing pulse edges must be equal to zero, because the EM fields should, at least, be continuous at any locations and any times within a uniform medium (even if the medium had dispersion).  As implicitly shown in the calculations by Mansuripur, the pulse edges, which meet the ``moving boundary condition'', will not produce additional Lorentz forces in the sense of time average [see Eq.\ (10) of Ref. \cite{r12} by setting $\phi_0=0$  and $T=$ an integer of wave periods].  In other words, the momentum transfer from the light pulse to the box only takes place on the vacuum-medium interface, while the pulse edges located inside the uniform medium do not have any contributions to momentum transfer.

\section{Photon's energy-associated mass and \\ momentum-associated mass \\ in a dielectric medium}
\label{appC}

\begin{figure*}
\begin{center}
\includegraphics[bb=1.25in 6.2in 7.25in 10.0in, scale=0.9]{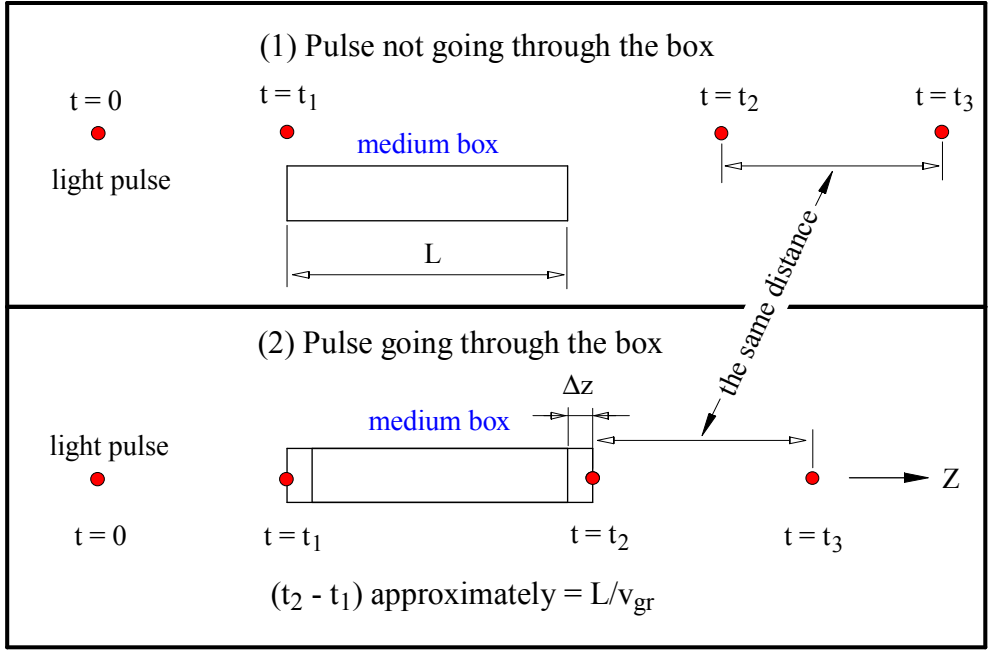}
\end{center}
\caption{Medium Einstein-box thought experiment for a short light pulse, analyzed by traditional center-of-mass-energy approach.  Case-1. Light pulse not going through the medium box.  Case-2. Light pulse going through the medium box: the pulse enters the box at  $t_1$, and when the pulse leaves the box at  $t_2$, the box shifts a distance of  $\Delta z$.  Both case-1 and case-2 have the same center of mass-energy. }
\label{fig2}
\end{figure*}

It is worthwhile to point out that, there are two kinds of mass: (i) energy-associated mass $m_{_E}$, defined through $E=m_{_E}c^2$  (Einstein's energy-mass equivalence formula), and (ii) momentum-associated mass  $m_{\mathbf{p}}$, defined through  $\mathbf{p}=m_{\mathbf{p}}\mathbf{v}$, where $E$,  $\mathbf{p}$, and $\mathbf{v}$ are, respectively, the particle energy, momentum, and velocity, with $P^{\mu}=(\mathbf{p},E/c)=(m_{\mathbf{p}}\mathbf{v},m_{_E} c)$ its four-momentum.  For classical particles and photons in vacuum, $m_{\mathbf{p}}=m_{_E}$ holds, while for photons in a medium, $m_{\mathbf{p}}=n^2_d\hbar\omega/c^2$  and $m_{_E}=\hbar\omega/c^2$  are valid, which lead to a Lorentz covariant Minkowski's four-momentum.  Thus we have  $P^{\mu}P_{\mu}=E^2/c^2-\mathbf{p}^2>0$ for classical massive particles, $P^{\mu}P_{\mu}=0$  for photons in vacuum, and $P^{\mu}P_{\mu}<0$  for photons in a medium.  Because of $m_{\mathbf{p}}\ne m_{_E}$  in a medium, the photon mass-\emph{vs}-momentum relation is different from that in vacuum where $m_{\mathbf{p}}=m_{_E}$  holds.  In other words, only $m_{\mathbf{p}}\mathbf{v}$  is the Lorentz covariant photon momentum in a medium, instead of $m_{_E}\mathbf{v}$.  

For an isolated system, the total momentum and energy are both conserved, namely $\sum \mathbf{p}_i=\sum m_{\mathbf{p}i}\mathbf{v}_i=const$  and $\sum E_i=\sum m_{Ei}c^2=const$, leading to the holding of $\sum m_{\mathbf{p}i}\mathbf{v}_i/\sum m_{Ei}=\mathbf{v}_c=const$.  Thus we have the mass-energy center $\mathbf{r}_c=\sum \int m_{\mathbf{p}i}d\mathbf{r}_i/\sum m_{Ei}$  moving uniformly.  Note that in $ \int m_{\mathbf{p}i}d\mathbf{r}_i$  the momentum-associated mass $m_{\mathbf{p}i}$  is involved, instead of the energy-associated mass  $m_{Ei}$.  To calculate $\mathbf{r}_c=\sum \int m_{\mathbf{p}i}d\mathbf{r}_i/\sum m_{Ei}$  in the dielectric Einstein-box thought experiment, $m_{\mathbf{p}}$  should be assumed to be known for the photon.  In the typical analysis by Barnett \cite{r7}, $m_{\mathbf{p}}$  for the photon in the medium is replaced by $m_{_E}=\hbar\omega/c^2$  (the same as that in vacuum).  However if $m_{\mathbf{p}}$  for the photon in the medium is known, then the photon momentum is actually known, equal to  $m_{\mathbf{p}}(c/n_d)$, with no further calculations needed, which is the straightforward way used by Leonhardt, except that he also uses $m_{_E}$  to replace $m_{\mathbf{p}}$ \cite{r20}.  From above, we can see that the Abraham's momentum in the dielectric Einstein-box thought experiment is derived actually by assuming an Abraham's momentum in advance.

To better understand why the Abraham's momentum is derived in the traditional analysis of the Einstein-box thought experiment \cite{r7}, specific illustrations are given below.

Figure \ref{fig2} shows the thought experiment consisting of a short light pulse and a transparent medium box.  Case-1 is for the pulse not going through the box, while case-2 is for the pulse going through the box.  The two cases have the same initial conditions and thus they have the same center of mass-energy.

\textbf{Case-1.}  Since the pulse does not go through the medium box and the box always keeps at rest, the mass-energy center for the system is given by
\begin{equation}
z_c=z_{c0}+\frac{m_{_E} ct}{(E+M_0 c^2)/c^2},~~~~~
\label{eqC1}
\end{equation}
where $z_{c0}$  is the initial mass-energy center, the pulse energy is $E=m_{_E} c^2$, and the box energy is  $M_0 c^2$.  In the vacuum, the pulse \emph{momentum}-associated mass and \emph{energy}-associated mass are the same, equal to  $m_{_E}$, and the pulse momentum is given by  $\mathbf{p}_{pulse-vac}=m_{_E} c\hspace{0.15em}\mathbf{\hat{z}}$.

\textbf{Case-2.}  The mass-energy center for the pulse going through the box is given by
\begin{equation}
z_c=z_{c0}~+~\frac{m_{_E} ct}{(E+M_0 c^2)/c^2}, ~~~~~(t<t_1)~~~~~~
\label{eqC2}
\end{equation}
\begin{align}
z_c=z_{c0}~+~&\frac{m_{_E} ct_1}{(E+M_0 c^2)/c^2}~+~\frac{m_{\mathbf{p}} v_{gr}(t-t_1)}{(E+M_0 c^2)/c^2}  \nonumber \\
~+~&\frac{p_{_M} (t-t_1)}{(E+M_0 c^2)/c^2}, ~~~~~( t_1\leq t <t_2)
\label{eqC3}
\end{align}
\begin{align}
z_c=z_{c0}~+~&\frac{m_{_E} ct_1}{(E+M_0 c^2)/c^2}  \nonumber \\
~+~&\frac{m_{\mathbf{p}} v_{gr}(t_2-t_1)}{(E+M_0 c^2)/c^2}~+~\frac{p_{_M} (t_2-t_1)}{(E+M_0 c^2)/c^2}~~  \nonumber \\
~+~&\frac{m_{_E} c(t-t_2)}{(E+M_0 c^2)/c^2}, ~~~~~(t\geq t_2)
\label{eqC4}
\end{align} \\
\noindent where $m_{\mathbf{p}}$  is the pulse momentum-associated mass in the medium, thus leading to the pulse momentum given by $\mathbf{p}_{pulse-med}=m_{\mathbf{p}}v_{gr}\mathbf{\hat{z}}$  with $v_{gr}$  the pulse energy velocity; the pulse energy is $E=m_{_E} c^2$, the same as in the vacuum; $p_{_M}$  is the box momentum, with $p_{_M}/M_0$  the box moving velocity, when the pulse is within the medium box.

When $t>t_1$  for the case-2, the pulse has entered the box, or just left the box, or has left the box and goes forward an additional distance, as shown in Fig.\ \ref{fig2}.   Comparing Eq.\ (\ref{eqC1}) with Eq.\ (\ref{eqC3}) or Eq.\ (\ref{eqC4}) we obtain the same equation of conservation of momentum, given by
\begin{align}
&p_{_M}=m_{_E} c-m_{\mathbf{p}}v_{gr}, ~~\mbox{or} \nonumber \\
&p_{_M}+m_{\mathbf{p}}v_{gr}=m_{_E} c.
\label{eqC5}
\end{align}
Namely, the sum of the momentums of the medium box and the light pulse, when the pulse enters the box, is equal to the momentum of the pulse in vacuum. 

When the pulse just leaves the box, the box has moved a distance, as shown in Fig.\ \ref{fig2}, given by
\begin{align}
\Delta z&=\frac{p_{_M}}{M_0}(t_2-t_1) ~~~~~~~~~~~~~~~~~~~~~~~~~~~~~~~~~~~~~ \nonumber \\
&=\frac{(m_{_E}c-m_{\mathbf{p}}v_{gr})}{M_0}\frac{L+\Delta z}{v_{gr}} \nonumber \\ 
&\approx\frac{(m_{_E}c-m_{\mathbf{p}}v_{gr})}{M_0}\frac{L}{v_{gr}} \nonumber \\ 
&=\frac{L m_{_E}}{M_0}\left( \frac{c}{v_{gr}}-\frac{m_{\mathbf{p}}}{m_{_E}} \right),
\label{eqC6}
\end{align}
where Eq.\ (\ref{eqC5}) and $(t_2-t_1)=(L+\Delta z)/v_{gr}\approx L/v_{gr}$  are used.

From above Eq.\ (\ref{eqC6}) we can see that, to obtain $\Delta z$  we have to assume that $m_{\mathbf{p}}$  is known.  In the traditional analysis, $m_{\mathbf{p}}=m_{_E}=E/c^2$  ($=\hbar\omega/c^2$ for a photon) is assumed, which leads to \cite{r7,r10}
\begin{equation}
\Delta z \approx\frac{L m_{_E}}{M_0}\left( \frac{c}{v_{gr}}-1 \right)=\frac{LE}{M_0c^2}\left( \frac{c}{v_{gr}}-1 \right).
\label{eqC7}
\end{equation}

However, as mentioned before, if $m_{\mathbf{p}}$  is known, then the pulse momentum $\mathbf{p}_{pulse-med}$  is actually known, equal to  $m_{\mathbf{p}}v_{gr}\mathbf{\hat{z}}$, without any further calculations.  Since $m_{\mathbf{p}}=m_{_E}=E/c^2$  is taken in the traditional analysis, $\mathbf{p}_{pulse-med}=m_{\mathbf{p}}v_{gr}\mathbf{\hat{z}}=(E/c)(v_{gr}/c)\mathbf{\hat{z}}$ is Abraham's momentum \cite{r10}.

For the single photon-medium box thought experiment, we have the Abraham photon momentum $m_{\mathbf{p}}v_{gr}=(\hbar\omega/c^2)(c/n_d)=\hbar\omega/(cn_d)$  if $m_{\mathbf{p}}=m_{_E}=\hbar\omega/c^2$  is taken \cite{r7}, while we have the Minkowski photon momentum $m_{\mathbf{p}}v_{gr}=(n^2_d\hbar\omega/c^2)(c/n_d)=n_d\hbar\omega/c$  if $m_{\mathbf{p}}=n^2_d\hbar\omega/c^2\ne m_{_E}$  is set.


\end{document}